\documentclass{pnastwo}

\usepackage{graphicx}
\usepackage{pnastwoF}
\usepackage{amssymb,amsfonts,amsmath,wasysym,cancel}

\contributor{Submitted to Proceedings of the National Academy of Sciences of the United States of America}
\url{www.pnas.org/cgi/doi/10.1073/pnas.0709640104}
\copyrightyear{2008}
\issuedate{Issue Date}
\volume{Volume}
\issuenumber{Issue Number}

\begin{document}

\title{Nonlinear conduction via solitons in a topological mechanical insulator}

\author{
 Bryan Gin-ge Chen\affil{1}{Instituut-Lorentz for Theoretical Physics, Leiden University, NL 2333 CA Leiden, The Netherlands},
 Nitin Upadhyaya\affil{1}{}
 \and
 Vincenzo Vitelli\affil{1}{}
}

\contributor{Submitted to Proceedings of the National Academy of Sciences of the United States of America}

\maketitle

\begin{article}

\begin{abstract}

Networks of rigid bars connected by joints, termed linkages, provide a minimal framework to design robotic arms and mechanical metamaterials built out of folding components. Here, we investigate a chain-like linkage that, according to linear elasticity, behaves like a topological mechanical insulator whose zero-energy modes are localized at the edge. Simple experiments we performed using prototypes of the chain vividly illustrate how the soft motion, initially localized at the edge, can in fact propagate unobstructed all the way to the opposite end. We demonstrate using real prototypes, simulations and analytical models that the chain is a mechanical conductor, whose carriers are nonlinear solitary waves, not captured within linear elasticity. Indeed, the linkage prototype can be regarded as the simplest example of a topological metamaterial whose protected mechanical excitations are solitons, moving domain walls between distinct topological mechanical phases. More practically, we have built a topologically protected mechanism that can perform basic tasks such as transporting a mechanical state from one location to another. Our work paves the way towards adopting the principle of topological robustness in the design of robots assembled from activated linkages as well as in the fabrication of complex molecular nanostructures.  


\end{abstract}


{\it Significance statement: 
Mechanisms are zero-energy motions that are key to the operation of
mechanical devices, from windshield wipers to robotic arms. We built
and studied mechanisms that are topologically protected: as in quantum
topologically protected systems, they are not affected by smooth
changes in material parameters. These prototypes are examples of
mechanical structures that we dub topological metamaterials. Their
excitations are topologically protected and yet tunable by changing
the geometry of the unit cell. We demonstrate that the continuum
elasticity of such mechanisms must necessarily address the
nonlinearity of the excitations that are solitons separating distinct
topological phases.
}

\begin{figure}
  \centering
  \includegraphics[width=.46\textwidth]{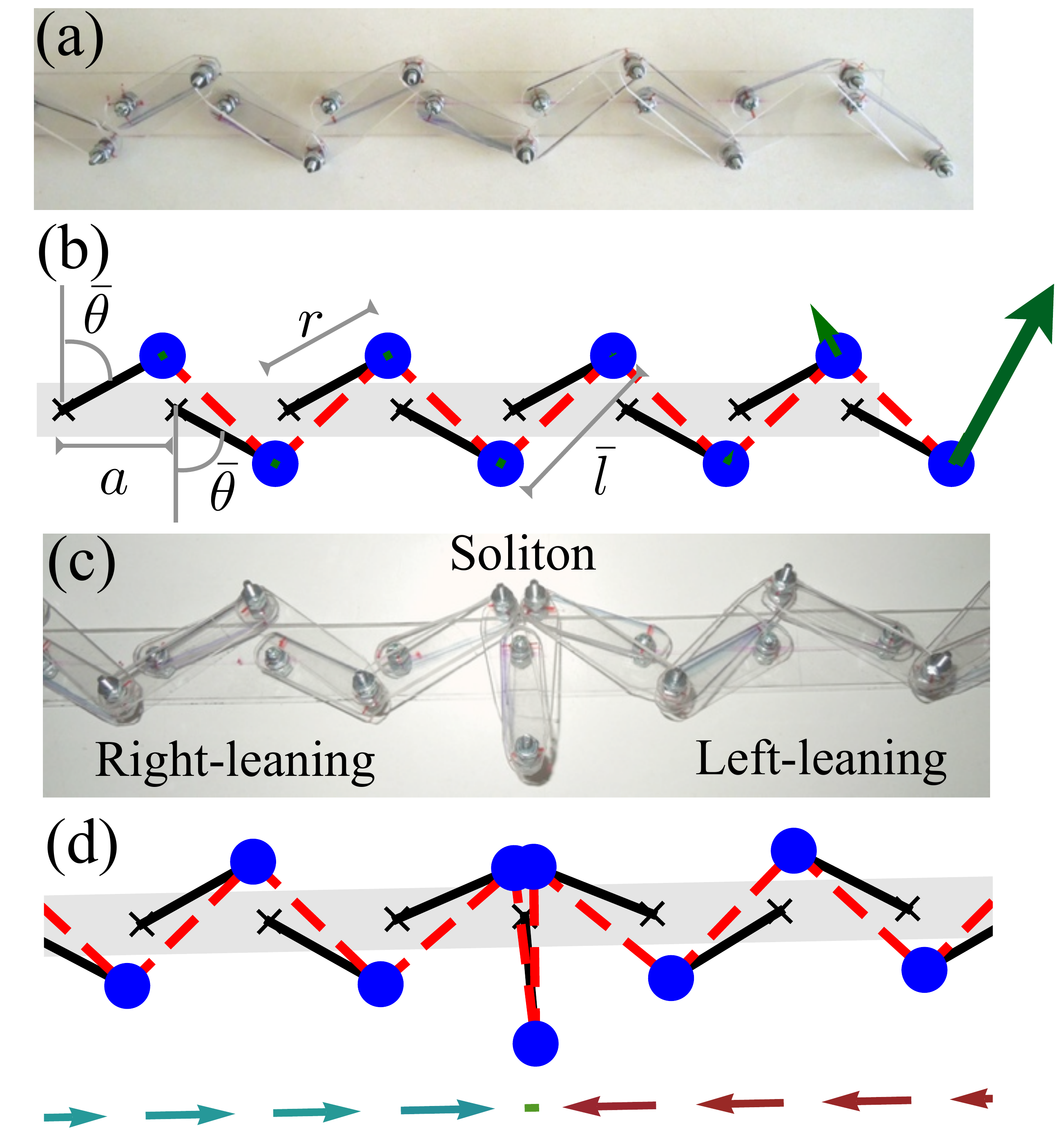}
  \caption{\label{fig:fig1}The chain of rotors in the flipper phase. (a) the
translation symmetric system with
$\theta=\bar\theta$ constant. We show a {\it linkage} made from plastic and metal
screws. (b) A computer sketch of the {\it elastic} chain
\cite{kanelubensky}: the masses are blue, rigid rotors are
black, and springs are dashed red lines. The green arrows depict
the amplitude of displacement of each mass of the edge-localized zero mode 
of the system. (c) A configuration of the linkage showing
a soliton as a domain wall between right-leaning and left-leaning
states. (d) A computer-simulated static configuration. The arrows beneath show the $x$-projections of each rotor.}
\end{figure}

\medskip
Mechanical structures comprised of folding components, such as bars or
plates rotating around pivots or hinges, are ubiquitous in
engineering, materials science and biology \cite{demaine}. For example, complex origami-like structures can be created by folding a paper sheet along
suitably chosen creases around which two nearby faces can freely rotate \cite{Maha,chris,guest}. Similarly, linkages can be viewed as
1D versions of origami where rigid bars (links) are joined at their ends by joints (vertices) that permit full rotation of
the bars, see Fig.\ (\ref{fig:fig1}a-c).  Some of the joints can be pinned to the plane while the remaining ones rotate relative to each
other under the constraints imposed by the network structure of the linkage \cite{thorpe}. Familiar examples include the windshield wiper,
robotic arms, biological linkages in the jaw and knee as well as toys like the Jacob's ladder \cite{edge} and the Hoberman sphere. Moreover, linkages and origami can be employed in the design of microscopic and structural metamaterials whose peculiar properties are controlled by the geometry of the unit cell \cite{katia,milton}.  

Many of these examples are instances of what mechanical engineers call {\it mechanisms}: structures where the degrees of freedom are nearly balanced by carefully chosen constraints so that the allowed free motions encode a desired mechanical function. However, as 
the number of components increases, more can go wrong: lack of precision machining or undesired perturbations. Robustness in this sense is a concern relevant to the design of complex mechanical structures from the microscopic to architectural scale, typically addressed at the cost of higher manufacturing tolerances or active feedback.

Here, we take an alternative approach inspired by recent
developments in the design of fault tolerant quantum devices
\cite{vincenzocommentary}. 
Consider, as an example, the quantized Hall conductivity of a 2D
electron gas that is topologically protected in the sense that it cannot change when the Hamiltonian is smoothly varied \cite{kanecolloquium}.
In this article, we present a topologically protected {\it classical} mechanism that can transport a mechanical state across a chain-like linkage without being affected by changes in
material parameters or smooth deformations of the underlying
structure, very much like its quantum counterparts. 

Kane and Lubensky \cite{kanelubensky} recently took an important step
towards establishing a dictionary between the quantum and classical
problems. Their starting point, that seems at first disconnected from
the linkages we study here, was to analyze the phonons in elastic systems composed of stretchable springs. In particular they derived a mathematical mapping between
electronic states in topological insulators and superconductors \cite{kanecolloquium}
and the mechanical zero modes in certain elastic lattices \cite{kaipnas}. The simplest is the 1D elastic chain, shown in Fig.\ (\ref{fig:fig1}b), inspired by the Su-Schrieffer-Heeger (SSH) model for polyacetylene \cite{ssh}, a linear polymer
chain with topologically protected electronic states at its free
boundaries. In the mechanical chain, the electronic modes map onto zero-energy vibrational modes with nontrivial topological index, whose eigenvectors represented as green arrows in Fig.\ (\ref{fig:fig1}b), are localized at one of the edges \cite{kanelubensky}. 
An intriguing question then arises: could these zero-energy edge modes propagate through the system in the form of finite deformations?

We address this question by building and analyzing a linkage of rigid
bars as an extreme limit of the 1D lattice of springs. This 
linkage allows no stretching deformations, yet it still
displays the distinctive zero-energy mode localized at the edges, see
Fig.\ (\ref{fig:fig1}) and Movie S1. By nudging the rotors along the
direction of the zero-energy mode (see Fig.\ (\ref{fig:fig1}b) and
Movie S2), we provide a vivid demonstration of how the initially
localized edge mode, can indeed propagate and be moved around the
chain at an arbitrarily small energy cost. We then show analytically
and numerically that the mechanism underlying the mechanical
conduction is in fact an evolution of the edge mode into a nonlinear
topological soliton, which is the only mode of propagation in the
chain of linkages that costs zero potential energy.  The soliton or
domain wall interpolates between two distinct topological mechanical
phases of the chain and derives its robustness from the presence of a
band gap within linear elasticity and the boundary conditions imposed
at the edges of the chain. Although the topological protection ensures
the existence of a domain wall, the dynamical nature of the soliton
falls into two distinct classes that can ultimately be traced to the
geometry of the unit cell. The prototypes we built therefore provide
simple examples of structures that we dub {\it topological
metamaterials} whose excitations are topologically protected
zero-energy solitons \cite{vincenzocommentary}.

\section{Topological band theory of phonons} 

The application of topological band theory to mechanics is most easily demonstrated in the context of the
1D elastic chain \cite{kanelubensky}, see
Fig.\ (\ref{fig:fig1}b). The model consists of a periodic
arrangement of alternating massless rigid rotors of length $r$ (black
bars), constrained to rotate about fixed pivot points (black
crosses), around an equilibrium angle
$\bar{\theta}$ at odd numbered sites and $\pi-\bar{\theta}$ at even
numbered sites. Here, the blue dots
denote point masses $M$, $a$ is the lattice spacing and
$\bar{l}=\sqrt{a^2+4r^2\cos^2\bar\theta}$ is
the equilibrium (unstretched) length of the Hookean spring (with
spring constant $k_e$) connecting the masses. Since the 1D mechanical system has $N_s=N$ lattice points and $N_b=N-1$ bonds, there is (at least) one zero-energy vibrational mode as required by constraint counting, i.e., $N_0=dN_s-N_b$, where $N_0$ is the number of zero modes and $d$ is the number of spatial dimensions. 

The rigidity matrix \cite{demaine,calladine} for the lattice in Fig.\ (\ref{fig:fig1})
\footnote{The authors of ref.\ \cite{kanelubensky} work with a matrix
$Q=R^T$.}, denoted $R$, is obtained by linearizing the change in length of
the spring that connects masses $\{n,n+1\}$ in terms of small angular
displacements $\delta\theta _{n,n+1}$ from the equilibrium value
$\bar{\theta}$, i.e., $\delta l_{n,n+1}=R_{ni}\delta\theta _{i}$. The
phonon modes are readily obtained from the Fourier-transformed
rigidity matrix $R(k)$ (considering the chain as a periodic
arrangement of two-particle unit cells)
\begin{align}
R(k)&=\begin{pmatrix}
q_+ & q_-\\
q_- & q_+e^{ik(2a)}
\end{pmatrix},
\end{align}
where
$q_\pm=\frac{r\cos\bar{\theta}(2r\sin\bar{\theta}\pm a)}{\sqrt{a^2+4r^2\cos^2\bar{\theta}}}$.
In this notation, the (Fourier-transformed) dynamical matrix reads $D(k)=R^\dagger(k)R(k)$. The eigenvalues of $D$ are the squared vibrational frequencies (in units of $\sqrt{k_e/M}$) of the normal modes of the lattice, given by $\omega(k)=|q_+\pm q_-e^{ika}|$, where the $\pm$ belongs to the acoustic (+) and optical branches (-) of the dispersion curve, respectively. 

For small $k$ (and $q_{+}>0,q_{-}<0$), the vibrational spectrum
of the acoustic branch is gapped, and $\omega^2(k)=\omega^2 _0 + c^2k^2$ where
\begin{align}
\omega _0 =&\left| \frac{r^2}{\bar{l}}\sin(2\bar{\theta})\right|>0
\end{align}
is the gap frequency and 
\begin{align}
c^2=&\frac{\cos^2\bar{\theta}|4r^2\sin^2\bar{\theta}-a^2|}{\bar{l}^2}\frac{a^2k_e}{M} \label{sound}
\end{align}
is the squared speed of sound. Owing to the gap, no propagating
modes (with real $k$) can be excited for $\omega < \omega _0$. Nevertheless, there is
a zero-energy mode, which corresponds to the specific value of
$k=\sqrt{-\omega _0}/c$ for which $\omega=0$, i.e., the value of $k$ for
which there are no changes in the spring length $\delta l _{n,n+1}=0$
for all $n$. Since $k$ is complex, the zero-energy mode is
non-propagating and thus Im $k$ yields an inverse penetration depth
$\ell^{-1}$ that evaluates to
\begin{align}
\frac{\ell}{a}
=&-\left[\ln\left|\frac{2r\sin\bar\theta-a}{2r\sin\bar\theta+a}\right|\right]^{-1}
\sim \frac{1}{\sin\bar{\theta}}.
\label{eqn:penetration} 
\end{align}
where the last relation is valid for small $\bar{\theta}$.
As shown in Fig.\ (\ref{fig:fig1}b), the eigenvectors $\vec{e}_i$ of
the dynamical matrix (shown as green arrows) have appreciable
magnitude only on the rightmost two particles illustrating the
exponential localization of the zero-energy
mode at the boundary.

Eq.\ (\ref{eqn:penetration}) shows that for $\bar\theta=\{0,\pi\}$;
$\ell$ diverges and the zero-energy end mode becomes an ordinary
infinite-wavelength acoustic phonon -- the chain is no longer gapped.
At $\bar\theta=\{\pi/2,3\pi/2\}$ the phonon spectrum collapses entirely to
0.  This demonstrates the topological robustness of the zero-energy edge modes: unless the chain is prepared with the specific
values of $\bar{\theta}=\{0,\pi/2,\dots\}$ for which the gap closes, their
presence is insensitive to changes in material parameters.

Whether the zero-energy mode is localized at the right or left edge is
determined by the topological polarization introduced in
\cite{kanelubensky}; here, simply the winding number
of 
the complex phase of $\det R(k)$. As $k$ goes from
$-\pi/(2a)$ to $\pi/(2a)$, the path of $\det R(k)$ in the complex
plane is a circle centered on the real axis at $q_+^2$ with radius
$q_-^2$, provided that $\bar{\theta} \ne 0$ \footnote{If
$\bar\theta=0$ the path of $\det R(k)$ passes through 0, making the
phase undefined.}. Thus, the winding number $n$ is 1 if $|q_+|<|q_-|$ and zero
if $|q_+|>|q_-|$ indicating that the mode is localized respectively to
the left ($\ell >0$ in Eq.\ (\ref{eqn:penetration})) or right ($\ell <0$)
edge of the chain.  The physical meaning of this classification is
apparent when considering the symmetry classes of the uniform ground
states: the black rotors in Fig.\ (\ref{fig:fig1}) can be either left-
($n=1$) or right-leaning ($n=0$).

Since the zero potential energy motion does not involve
stretching or compression of the spring, it can be studied both in the
hybrid spring-strut system introduced by Kane and Lubensky as well as
in the chain of linkages shown in Fig.\ (\ref{fig:fig1}a) and
(\ref{fig:fig1}c). In these prototypes, the plastic rotors rotate
around bolts attached to a longer piece of plastic that
serves as the rigid background, and are attached at their ends by
other plastic pieces. Self-intersections are avoided by arranging 
alternating bars at different heights in the transverse direction. 

\section{Beyond phonons: solitons in systems of linkages}

The linear elastic theory reviewed in the previous section predicts
that there are no bulk low-energy phonons below the gap frequency and
that the zero-energy mode is exponentially localized at the edge --
i.e., the chain is a topological
{\em mechanical insulator} (Movie S1). However, this conclusion 
is manifestly at odds with the simple experiments we
performed using prototypes of the SSH chain with rigid constraints, as
shown in Fig.\ (\ref{fig:fig1}). 

By tilting the chain, a soft mode
initially localized at the edge of the chain propagates under the
effect of gravity all the way to
the opposite end, as shown vividly in Movie S2. This simple
experiment demonstrates that the chain is in fact a {\em mechanical
conductor} whose carriers are nonlinear solitary wave excitations not
captured within linear elasticity.  The nonlinear mechanism of
conduction is the central result of our work and relies on a key
difference between vibrational and electronic states: 
phonons are infinitesimal physical displacements that can be
integrated to finite deformations of the underlying mechanical
structure, whereas electronic states live in an abstract Hilbert
space. 

The SSH chain is a paradigmatic and
analytically tractable representative from a broader class of floppy
mechanical systems that share a common feature -- the {\it
infinitesimal} zero-energy motion described by the localized
topological modes extends to a {\it finite} zero-energy
motion that propagates freely through the bulk. We expect similar
behavior to occur in 2D and 3D structures involving
linkages or other general mechanical structures that could be of
considerable interest for robotic manipulations and design of
meta-materials \cite{hawke}.

\begin{figure}
  \centering
  \includegraphics[width=.44\textwidth]{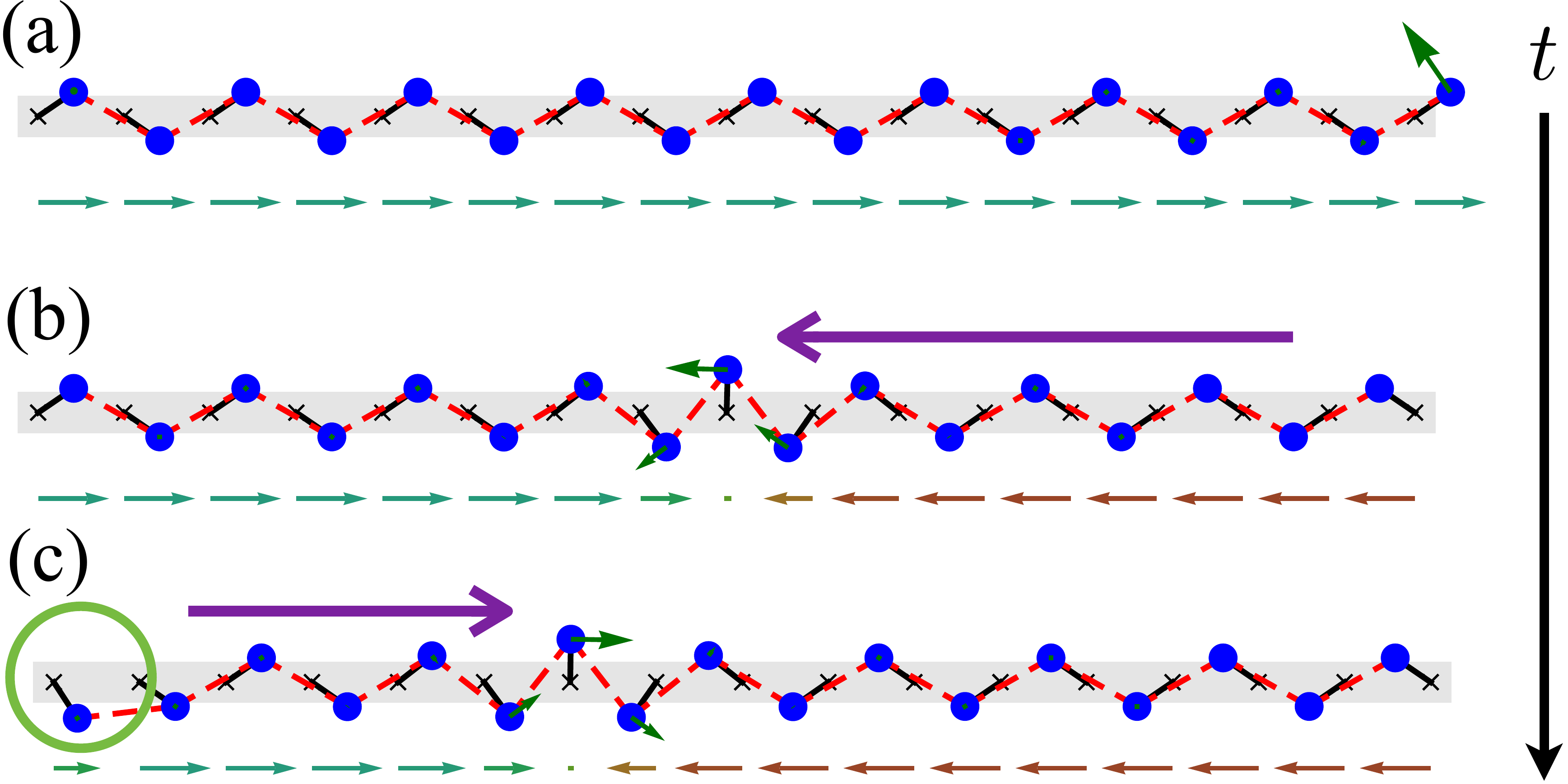}
  \caption{\label{fig:fig2} The evolution of a
flipper soliton arising from integrating the zero mode of a finite
chain ($r/a=0.5,\bar\theta=0.97$), with $x-$projections of the rotors
underneath each snapshot (see also Movies S3 and S4). The system
evolves from (a) right-leaning to (b) left-leaning and
then back to (c) right-leaning.  The green arrows attached to the rotors
depict the amplitude
of the zero mode on each mass; note that it is always localized to the domain wall. In the flipper phase, the
angle of a rotor flips between $+\bar\theta$ and $-\bar\theta$ each
time the soliton passes through, except for the rotors on the ends,
which also toggle between up-pointing and
down-pointing. Hence in a full cycle the soliton goes from right to
left and back {\em twice}.}
\end{figure}

What types of nonlinear waves does this type of system support? Are there
different phases of motion?  To answer these questions we
performed Newtonian dynamics simulations of the chain for different
values of the geometrical parameters $\{\bar l,r,a\}$ and the spring
constant, $k_e$. Newton's laws are numerically integrated with the
initial condition set by the zero-energy mode: 
$\vec{v}_i^0=v_0 \vec{e}_i$, where $\vec{v}_i^0$ is the initial
velocity of the $i$th particle, $\vec{e}_i$ is the 
component corresponding to the $i^{th}$ particle in the 
zero-energy eigenvector of the dynamical matrix, and $v_{0}$ is a factor determining the initial speed.
As seen in Fig.\ (\ref{fig:fig2}) and Movie S3, since the
zero-energy mode is initially localized at the edge, the above
procedure amounts to imparting an initial kick to the first few
particles. We discuss below how this initial impulse evolves
and propagates through the chain as well as the rich
phenomenology that emerges as we vary the geometrical parameters of
the chain.

In Fig.\ (\ref{fig:fig2}a), we show a chain with the rotors
initially in the right-leaning mechanical state. As the rotors at the right end
rotate following the zero-energy eigenmode (marked with small green
arrows) their angles soon reach $-\bar\theta$ and they enter
the left-leaning state and halt. However, rotors further to the left
now begin to rotate, generating a moving, localized region of motion.  At the
far left of the chain, the rotors remain in the right-leaning 
state until that region reaches them.  The dynamics thus generate a domain wall interpolating between the
right- and left-leaning states. 
Fig.\ (\ref{fig:fig2}b) shows a snapshot of the chain where the domain wall is halfway to the left end.  Since the passage of the
domain wall flips the direction of the rotors from $\theta=\bar\theta$
to $\theta=-\bar\theta$ (measured with respect to the positive
$y$-axis on the odd-numbered sites) we label this the {\em flipper}
phase of motion. Once the domain wall has reflected off the
left edge of the chain, the edge rotor now points down (Fig.\
(\ref{fig:fig2}c)). Only when the domain wall
has traversed the chain back and forth {\em twice} 
does the entire chain return to its initial right-leaning
state.  
Upon varying the geometrical parameters of the chain, in particular,
after increasing the ratio $r/a$, we find that the rotors
in the flipper can also overshoot their equilibrium positions and the
profile exhibits oscillations, though the final effect of the soliton is still to
flip the rotors between the two states.  We call this variant the {\em wobbling
flipper} (Movie S4).

\begin{figure}
  \centering
  \includegraphics[width=.44\textwidth]{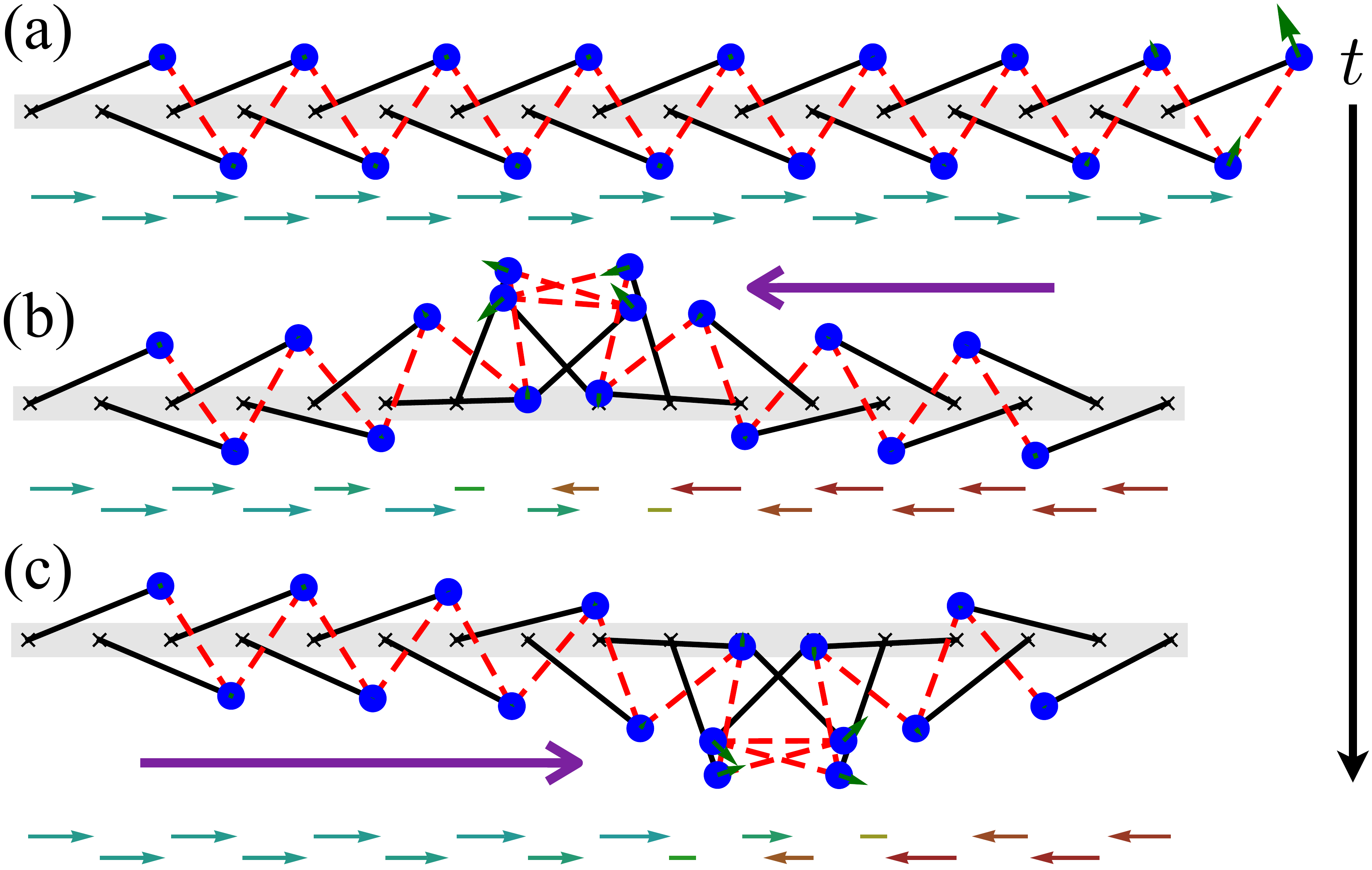}
  \caption{\label{fig:fig3} The
evolution of a spinner soliton arising from integrating the zero mode
of a finite chain ($r/a=2,\bar\theta=1.18$), with $x-$projections of
the rotors underneath each snapshot (see also Movie S5). The
system evolves from (a) right-leaning to
(b) left-leaning and then back to (c) right-leaning.
The green arrows attached to the rotors
depict the amplitude of the zero mode on each mass; note that it is
always localized to the
domain wall. In the spinner phase, the angle of each rotor advances by $+\pi$
each time the soliton passes.}
\end{figure}

After further increasing $r/a$, a strikingly different behaviour
emerges (Fig.\ (\ref{fig:fig3}) and Movie S5).
Although a domain wall interpolating between the right- and
left-leaning mechanical states is still observed (Fig.\
(\ref{fig:fig3}b)), the rotors now rotate counterclockwise by an angle of $\pi$ each time a domain wall passes by. 
We refer to this as the {\em spinner} phase of motion. As the domain wall
reflects off the ends of the chain, the edge rotors rotate by $2\pi$ and
complete a full circle. In contrast to the flipper phase, the
initial state of the chain is restored after the domain wall has
completed one cycle around the chain. 

\begin{figure}
  \centering
  \includegraphics[width=.44\textwidth]{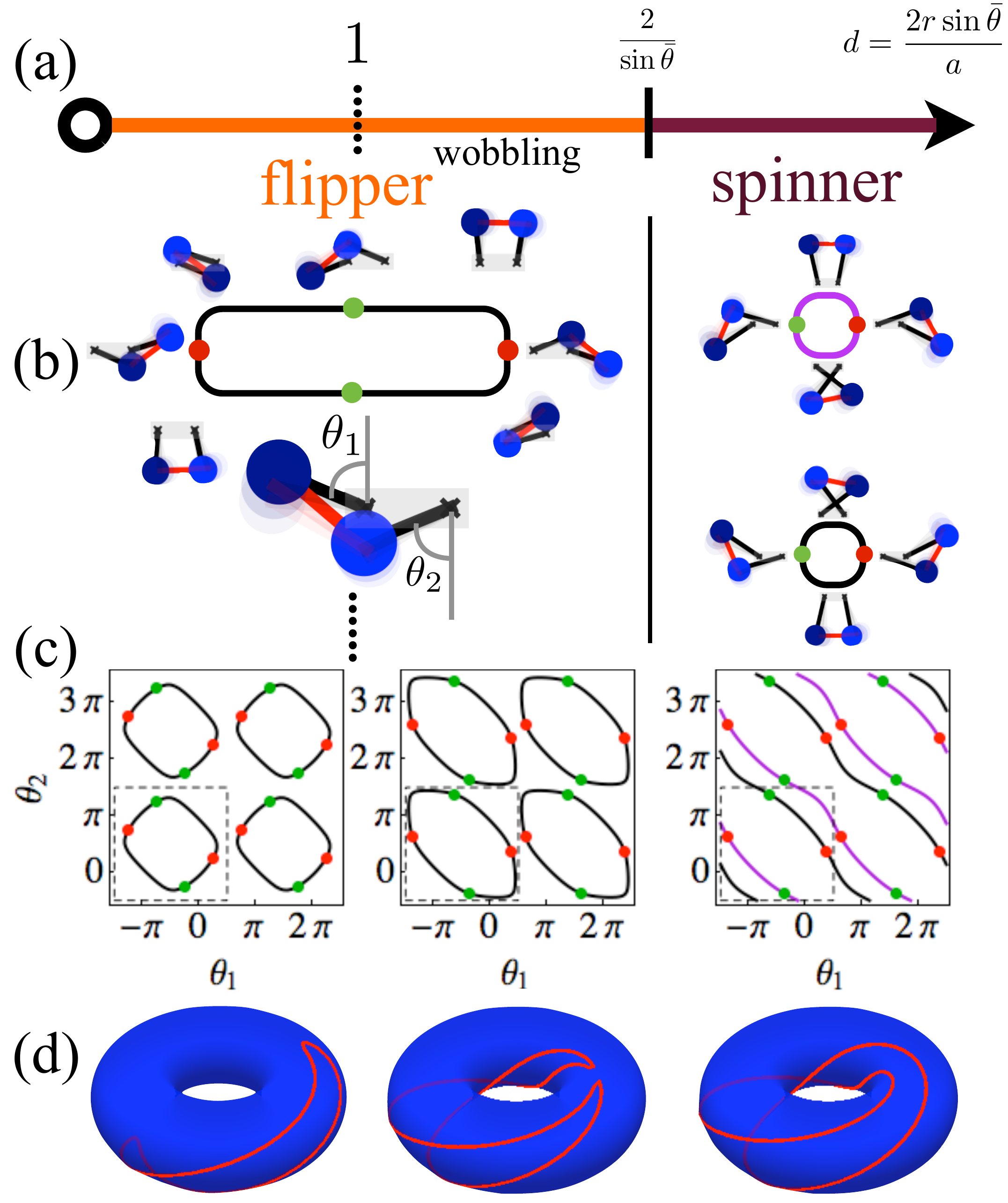} 
  \caption{\label{fig:fig4} (a) Phases of the soliton as a function of
$d=\frac{2r\sin\bar\theta}{a}$.  (b) The unit cell
(4-bar linkage) configuration spaces in the flipper and spinner
phases (four copies of each); the translation-symmetric points are marked in red
(right-leaning) and green (left-leaning). There are two connected
components in the spinner phase and just one in the flipper phase. (c)
Pictures of the
unit cell configuration space in the torus of $\theta_1,\theta_2$
angles (defined by 
$\delta l_{1,2}=0$). (d) The topology of the configuration spaces
depicted on 3D tori. In going from flipper to spinner, one circle splits into two linked Villarceau-like rings.}
\end{figure}

\section{Phase diagram of nonlinear excitations}

In order to understand the
transitions and differences between these phases of motion, we
can treat the chain as a 1D metamaterial and explain the nonlinear
dynamics in terms of the geometry of the unit cell, parametrized by
the dimensionless number $d=\frac{2r\sin\bar\theta}{a}$. 
Consider the zero-energy configuration space of the four-bar linkage
which is the unit cell of the chain. This space is defined by the zero-stretching
constraint $\delta l_{n,n+1}=0$ (for the spring connecting rotors $n$
and $n+1$) and hence we replace the spring by a strut in what follows. 
In Fig.\ (\ref{fig:fig4}b), we show snapshots of the unit cell
parametrized by the angles $(\theta _1,\theta _2)$, while in Fig.\
(\ref{fig:fig4}c), we show the corresponding configuration spaces for flippers (left),
wobblers (middle) and spinners (right) respectively (Movies S6, S7,
and S8 show full cycles of motion for the unit cells in all three
cases, alongside animations of the unit cell configuration
spaces). The red (right-leaning) and green (left-leaning) dots in Fig.\ (\ref{fig:fig4}b-c) represent four special configurations for which
$\theta=\pm\bar\theta,\pi\pm\bar\theta$ and correspond to the
four spatially periodic ground states that are related to each other by reflection symmetries.

In going from left to right in Fig.\ (\ref{fig:fig4}c) (increasing
$d$), the configuration space evolves from one
connected circle bounding a disk on a torus (parametrized by the coordinates
$(\theta _1,\theta _2)$) to two diagonal circles ``linked'' on that
torus. This happens precisely when $\bar{l}$ decreases below $2r-a$,
equivalently $d >2/\sin\bar\theta\equiv d_c$, thus marking the
transition from the flipper to the spinner phase. In Fig.\
(\ref{fig:fig4}d), we show sketches of the phase portrait on a
3D torus to emphasize this doubly-periodic configuration space. Thus, the
boundary that separates the flipper phase from the spinner phase is in
fact a topological change in the configuration space of the four-bar
linkage \cite{kapovichmillson}. 

To understand this change, observe that when $r \ll
a$ we can smoothly access all the possible configurations of a linkage
of four bars (Fig.\ (\ref{fig:fig4}b) left panel). However, as $r/a$
increases, the four-bar linkage approaches a triangle
that is pinned at one vertex. Since it is impossible
to transform a triangle to its mirror image via translations and
rotations in the plane, the set of configurations becomes disconnected
(Fig.\ (\ref{fig:fig4}b) right panel).
Thus, for $d>d_c$ (spinners) the unit cell
configuration space consists of two components, each containing a
single pair of ground states $\pm\bar\theta$ and $\pm\bar\theta+\pi$.
A period of the motion of the four-bar linkage reveals how the
soliton propagates through the chain -- all rotors rotate
in the same direction, from $\bar\theta$ to $\bar\theta+\pi$, say. By contrast, for
$d<d_c$ (flippers), a full cycle of the unit cell
visits all four ground state configurations. The transition between
flipper and spinner states appears to be discontinuous, as we show when we discuss the continuum description of the solutions.

To understand the transition between the non-wobbling flipper and the
wobbling flipper phases, notice that as a flipper soliton passes
through, the rotors must eventually rotate from $\bar\theta$ to $-\bar\theta$ passing through
$\theta=0$. Suppose the first and second masses at the edge initially both rotate
counterclockwise. By virtue of the flipper motion, the second mass must eventually
rotate clockwise, thus it will appear to wobble.
This is visible in Fig.\ (\ref{fig:fig4}c) as the change in sign of the slope
$d\theta_2/d\theta_1$ at all the red and green points, and the
threshold for this can be derived from the point at which the
penetration depth $\ell$ in Eq.\ (\ref{eqn:penetration}) vanishes:
$2r\sin\bar\theta=a$ or $d=1$.  

The existence of these rich phases of motion illustrate the fact that the uniform ground
states and localized zero-energy edge mode are best
viewed as snapshots of a periodic nonlinear motion and its velocity
field.

\section{Continuum theory: flipper solitons}

In this section and the next we discuss how the flipper and spinner motions discussed qualitatively
in the previous section emerge as topological soliton solutions to the very equation that within
the linear approximation predicts a localized edge mode: the
constraint equation $\delta{l_{n,n+1}}=0$. These solitons are
described by solutions to the $\phi^4$ and sine-Gordon equations, but have the key
additional feature that they cost precisely zero potential energy.

We begin by deriving the equation of motion of the flipper solitons in the limit $d\ll 1$.
In terms of the angles of the rotors in a unit cell $\theta
_n$ and $\pi-\theta _{n+1}$ (measured clockwise with respect to the
$+y-$axis), the constraint equation $l^2_{n,n+1}=\bar{l}^2$ reads,
\begin{eqnarray}
\cos(\theta _{n} + \theta _{n+1}) - \cos(2\bar{\theta}) + \frac{a}{r}(\sin\theta _{n} - \sin\theta _{n+1})=0. \label{flip_discrete}
\end{eqnarray}
To take the continuum limit of Eq.\ (\ref{flip_discrete}),
we define a slowly-varying angular field $\theta (x)$ by letting
$a\rightarrow dx$, $\theta
_n\rightarrow\theta(x)-\frac{a}{2}\frac{d\theta}{dx}$, $\theta
_{n+1} \rightarrow \pi-\theta(x) - \frac{a}{2}\frac{d\theta}{dx}$
(using a Taylor series expansion centered at $x=n+1/2$). We obtain
\begin{eqnarray}
\cos(2\bar{\theta}) - \cos(2\theta) - \frac{a^2}{r}\cos\theta\frac{d\theta}{dx}=0.\label{eq:firstorder}
\end{eqnarray}
Note that this choice preserves the underlying reflection symmetry
$(x,\theta)\rightarrow(-x,-\theta)$.

In terms of the field $u(x)=r\sin\theta (x)$ (the $x$-component of the
position of the mass), Eq.\ (\ref{eq:firstorder}) reads,
\begin{eqnarray}
\frac{a^2}{2}\frac{du}{dx} = u^2-\bar{u}^2, \label{phi4_1}
\end{eqnarray}
where $\bar u\equiv |r\sin\bar\theta|>0$. Besides the uniform left- and right-leaning solutions
$u(x)=\pm \bar u$, Eq.\ (\ref{phi4_1}) admits only one {\em zero-energy} solution (for $u<\bar{u}$) given by the kink 
\begin{equation}
u(x) = \bar u\tanh\left(\frac{x-x_0}{w_0}\right), \label{width}
\end{equation}
where $w_0=\frac{a^2}{2\bar u}=\frac{a}{d}$ is the width of the
static domain wall that interpolates between
left-leaning ($u<0$ as $x\rightarrow-\infty$) and right-leaning
($u>0$ as $x\rightarrow+\infty$) states. Note that $w_0$ is
proportional to the penetration length $\ell$ derived in Eq.\
(\ref{eqn:penetration}) for small $\bar\theta$ and diverges when the
gap closes. This is analogous to the divergence of domain wall widths
at the critical point in the Landau theory of second order phase
transitions. 

\begin{figure}
 \centering
 \includegraphics[width=.44\textwidth]{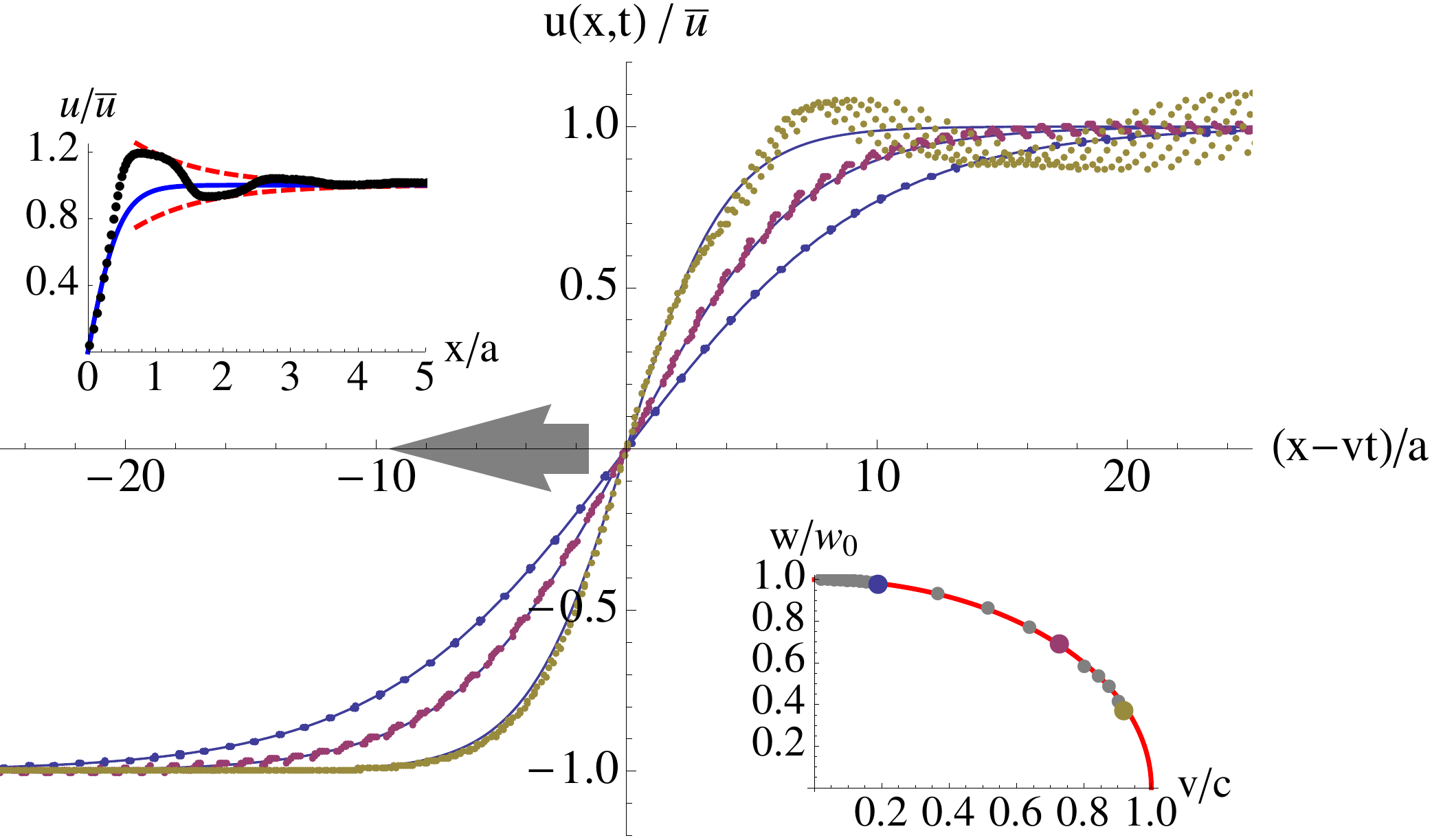}
 \caption{\label{fig:flippercompare} A comparison of the numerically
generated displacement field $u(x-vt)$ (symbols)  with the continuum
profiles in Eq.\ (\ref{Kink2}) (solid lines) in the flipper phase. The
numerical results are for a chain of 150 rotors with
$\bar\theta=0.1,r/a=0.5$ and $k_e=1,M=1$. Colours denote different
propagation speeds, increasing from blue to gold dots.  The speed of
propagation $v$ used in Eq.\ (\ref{Kink2}) was measured
 by tracking the motion of the center of mass of the
domain wall (obtained by interpolating the
data points $(j,\theta_j)$ and choosing the value of $x$ where the
line passes through $\theta=0$).  The curves corresponds to 10 time
snapshots of a single chain, each translated so that the center of the
soliton is at $x-vt=0$. Bottom inset: the numerically extracted width
of the kinks (data points obtained from the inverse of the slope of
$\operatorname{arctanh}
u/\bar u$ for the recentered profiles) is compared with the Lorentz
contraction factor (red line).  Top
inset: data from a wobbling flipper profile with
$\bar\theta=0.77,r/a=1.45$ (black points) is compared with a $\tanh$
profile (blue) with the flipper width and an exponential decay
envelope (red) with the spinner width. See SI Appendix and Figs. S1-S2
for a more detailed look at the wobbling flipper profiles.}
\end{figure}

In the SI Appendix, we derive the continuum Lagrangian for the chain in the flipper phase in the limit $l(x)\approx\bar{l}$ and $\bar{\theta}\ll1$,
which reads:
\begin{align}
\mathcal{L}=&\int dx \left[\frac{1}{2}M\left(\frac{\partial
u}{\partial t}\right)^2 -
\frac{1}{2}K\frac{a^4}{4}\left(\frac{\partial u}{\partial
x}\right)^2\right.\nonumber \\
&\left.-\frac{1}{2}K(\bar u^2-u^2)^2 - \frac{1}{2} K \frac{a^2}{2}(\bar
u^2-u^2)\frac{\partial u}{\partial x}\right].\label{lagrangian2}
\end{align}
In addition to the ordinary $\phi^4$ potential, $\mathcal{L}$ has an
additional boundary term linear in $\partial_x u$. This extra
term ensures that the static kink has zero energy since the last three
terms in Eq.\ (\ref{lagrangian2}) can be written as a perfect square
that vanishes for the static kink solution in Eq.\ (\ref{phi4_1})
\cite{BPS1,BPS2}. It also breaks the $\partial_x u \rightarrow
-\partial_x u$ symmetry of the ordinary $\phi^4$ theory. As a result,
the anti-kink solution of Eq.\ (\ref{lagrangian2}) (with left- and
right-leaning states reversed in space) costs a finite stretching
energy. Hence, the anti-kink is forbidden in the linkage limit, where
$k_e \rightarrow \infty$ and only solutions of Eq.\ (\ref{phi4_1}) are
admitted. We can define the topological charge $Q$
as the difference in the number of kinks and anti-kinks:
\begin{equation}
Q = \frac{1}{2 \bar{u}} \int_{-\infty}^{+\infty} \frac{\partial u}{\partial x} dx = \frac{1}{2 \bar{u}} \left[u(+\infty)-u(-\infty)\right]. 
\label{soliton_charge}
\end{equation}
For a system with periodic boundary conditions, $Q=0$, and no solitons
exist. For a system that has a left- and right-leaning edge, $Q=1$,
and the linkage must support one and only one kink that is therefore
topologically protected.  The topological index is thus a measure of
the number of solitons in the system, and is consistent with the
number of zero modes from constraint counting.

Since the $\phi^4$
theory is Lorentz invariant, the dynamical solution is simply obtained
by a Lorentz boost 
\begin{eqnarray}
u(x,t) = \bar u\tanh\left[\frac{x-x_0-vt}{\frac{a^2}{2\bar u}\sqrt{1-\frac{v^2}{c^2}}}\right],\label{Kink2}
\end{eqnarray}
where $v$ is the speed at which the kink propagates (set by the
initial kinetic energy in the system) and $c$ is the
linear speed of sound. From Eq.\ (\ref{sound}), the speed of linear
sound (defined from the linear part of the dispersion curves above the gap)
for $r\ll a$ is $c\approx \dfrac{a^2}{\bar{l}}\sqrt{\frac{k_e}{M}}$
valid for small $\bar{\theta}$.  In the comoving frame, Eq.\
(\ref{Kink2}) is equivalent to Eq.\ (\ref{width}), provided that the {\it static} width $w_0$ is replaced by the Lorentz contracted width $w=w_0 \sqrt{1- \frac{v^2}{c^2}}$. For linkages $c$ diverges, hence $w=w_0$.

To test our continuum approximation for the domain wall in the discrete mechanical chain, we numerically obtain the displacement
field $u(x,t)$ of the rotors for a range of initial energies imparted to the chain.
In Fig.\ (\ref{fig:flippercompare}), we show that the displacements (dotted curves) in the flipper phase compare
favourably with the continuum $\phi^4$ predictions in Eq.\
(\ref{Kink2}) (solid lines).  
The predicted Lorentz contraction is evident in the profiles in Fig.\ (\ref{fig:flippercompare}) as a decrease in the width of the profiles 
when the speed of propagation $v$ is increased (see bottom inset). Note the presence of a
large wake behind the moving domain wall, exhibited in the profile
with the highest energy (gold symbols). 
We leave a more detailed analysis of the wake and higher-order nonlinearities for future work. 

The wobbling flipper phase is more challenging to treat analytically.
As shown in the top inset of Fig.\ (\ref{fig:flippercompare}), the
profile exhibits oscillations around the values $\pm\bar\theta$.
Numerically, the slope at the kink center seems to follow the
prediction coming from the inverse width of the continuum flipper
solution (blue curve).
However, the oscillations in the profile decay exponentially with a decay length of
approximately $r\sin\bar\theta$ (red dashed envelope).  As we shall see in the next section,
this length scale is associated with the width of the spinner. In the
SI Appendix, we derive a linearized approximation for the wobbling
flipper which recovers the oscillations qualitatively and recovers
precisely this decay length (Figs. S1-S2), shedding light on how the transition
between flipper and spinner phases occurs. 

\begin{figure}
\centering
\includegraphics[width=.44\textwidth]{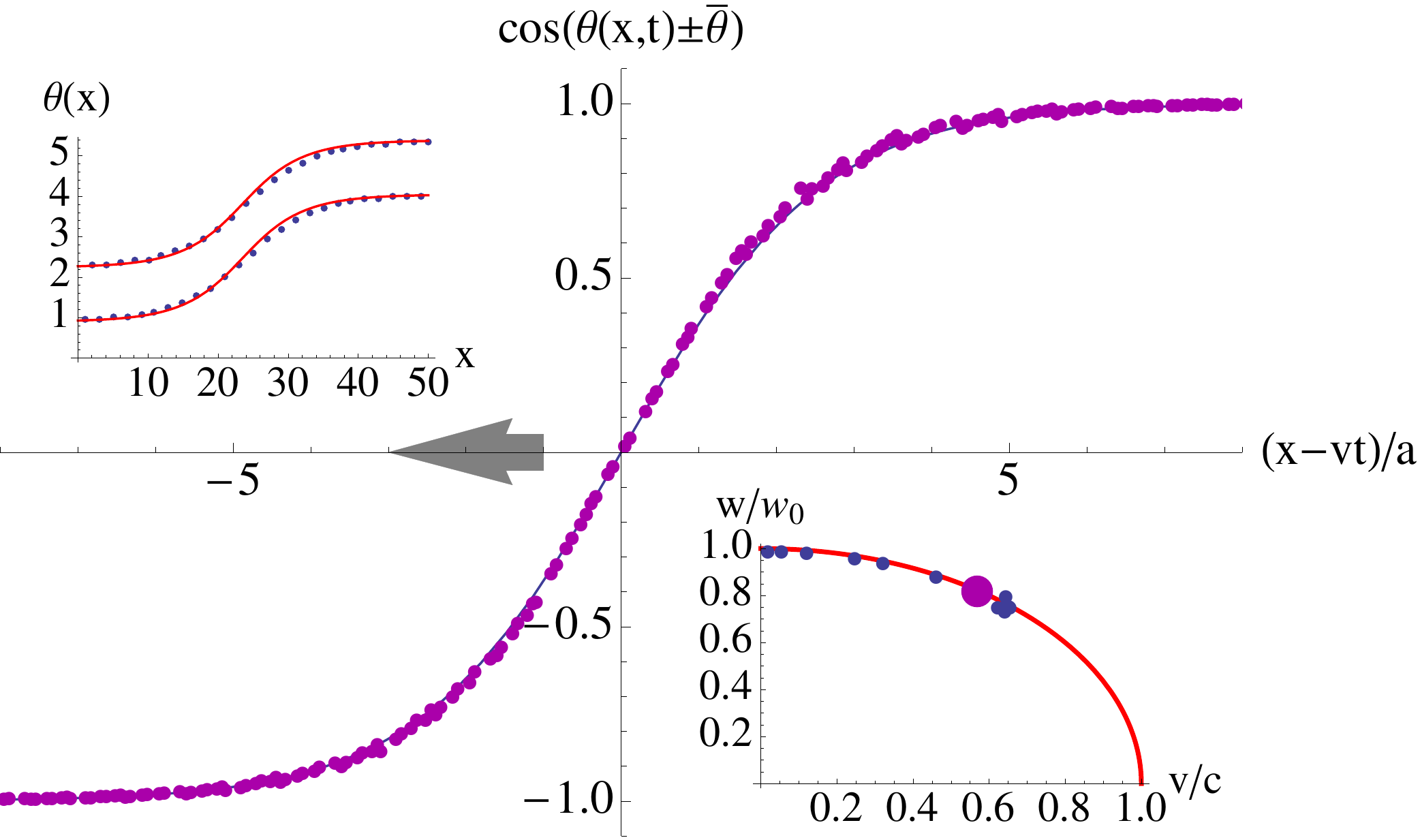}
\caption{\label{fig:spinnercompare}A comparison of the numerically
generated field $\cos(\theta(x-vt)\pm\bar\theta)$ (symbols) with the
continuum profiles in Eq.\ (\ref{sgd}) (solid line) in the spinner
phase. The numerical results are for a chain of 50 rotors with
$\bar\theta=0.9,r/a=4.0$ and $k_e=1,M=1$. Profiles and speeds were
obtained as in Fig.\ (\ref{fig:flippercompare}), except that the even
and odd sites are first shifted appropriately to yield continuous profiles. Top inset:
$\theta(x)$ for one snapshot showing how the even and odd angles are
offset by the constant $\pi-2\bar\theta$. Bottom inset: the
numerically extracted width of the kink (as in Fig.\
(\ref{fig:flippercompare})) is compared with the Lorentz contraction factor.}
\end{figure}

\section{Continuum theory: spinner solitons}

The solution presented for the flipper phase assumes that $u$ never
exceeds $\bar u$, a condition that is certainly violated by the
spinner solitons in which the rotors rotate by $\pi$. Moreover, in the
spinner phase where $r\gg a$, pairs of neighbouring rotors in a unit
cell seem to move nearly in phase like adjacent sides of a rigid
triangle (Movie S5).  In order to describe the motion in this phase, we
thus construct a description for the dynamics of
spinner solitons by integrating out the motion of every other rotor (say, the
even sites). 

Let $l_{n,n+1}$ be the length of the springs between nodes $\{n,n+1\}$. Then, the
constraint equation $l^2_{n,n+1}-\bar{l}^2_0=l^2_{n+1,n+2}-\bar{l}^2$
can again be expressed in terms of the geometrical parameters $r,a$ and the
angles $\{ \theta_n, \theta _{n+1}, \theta _{n+2}\}$ as before.  
We next ``integrate out'' $\theta_{n+1}$, by using the
constraint equation $l^2_{n,n+1}=\bar{l}^2$. Expressing this second
constraint equation in terms of the angles $\{ \bar{\theta}, \theta
_{n}, \theta _{n+1}\}$, we find (in the limit $a \ll r$) that
$\theta _{n+1} \approx \theta _{n} + \pi
-2\bar{\theta}$ \footnote{We explicitly tested the validity of this
approximate relation in classical dynamics simulations.}. 

Taking the continuum limit, 
$\theta _n\rightarrow
\theta (x)$ and $\theta _{n+2}\rightarrow \theta(x+2a)=\theta +
2a\theta'$ (where $f'\equiv df/dx$), and retaining terms to leading order in $a$, we obtain the following
differential equation for the {\it static} profile valid in the
spinner phase
\begin{eqnarray}
\theta' = -\frac{1}{r}\left(\frac{\sin(\theta-\bar{\theta})}{\sin\bar{\theta}}\right)
\end{eqnarray}
that can be easily integrated with the result 
\begin{eqnarray}
\cos(\theta-\bar{\theta}) = \tanh\left(\frac{x}{r\sin\bar{\theta}}\right).
\label{sg}
\end{eqnarray}
Thus, the static spinner profile at {\it alternate} sites is described by a sine-Gordon soliton of the form in Eq.\ (\ref{sg}) whose width is
$w_0=r\sin\bar{\theta}=ad/2$. This is in contrast with the flipper
phase of motion where the width of the soliton scales inversely with
$d$ (Eq.\ (\ref{width})). Indeed, the basic continuum
field in the spinner phase lives on only half the rotors, as compared
to the flipper phase where the field lives on all the rotors. Thus at
a very basic level, the transition between flippers and spinners
cannot be described as a smooth transition between two profile shapes. 

The dynamical solution is again obtained by a Lorentz boost
\begin{eqnarray}
\cos(\theta \pm\bar{\theta}) = \pm\tanh\left(\frac{x-vt}{r\sin(\bar{\theta})\sqrt{1-\frac{v^2}{c^2}}}\right),
\label{sgd}
\end{eqnarray}
where $\pm$ correspond to the solutions for even (odd) sites
respectively and from Eq.\ (\ref{sound}), the speed of sound in the
spinner phase ($r\gg a$) is 
$c= \dfrac{ar\sin(2\bar\theta)}{\bar{l}}\sqrt{\frac{k_e}{M}}$.
In the SI Appendix, we re-derive the sine-Gordon solution starting
from the discrete Lagrangian for our mechanical chain, valid in the
spinner phase. 
In Fig.\ (\ref{fig:spinnercompare}) we compare the continuum
sine-Gordon predictions (solid line)
to numerical solutions (symbols), and find good agreement in the
dynamical width as well (bottom inset).  


\begin{figure}
  \centering
  \includegraphics[width=.46\textwidth]{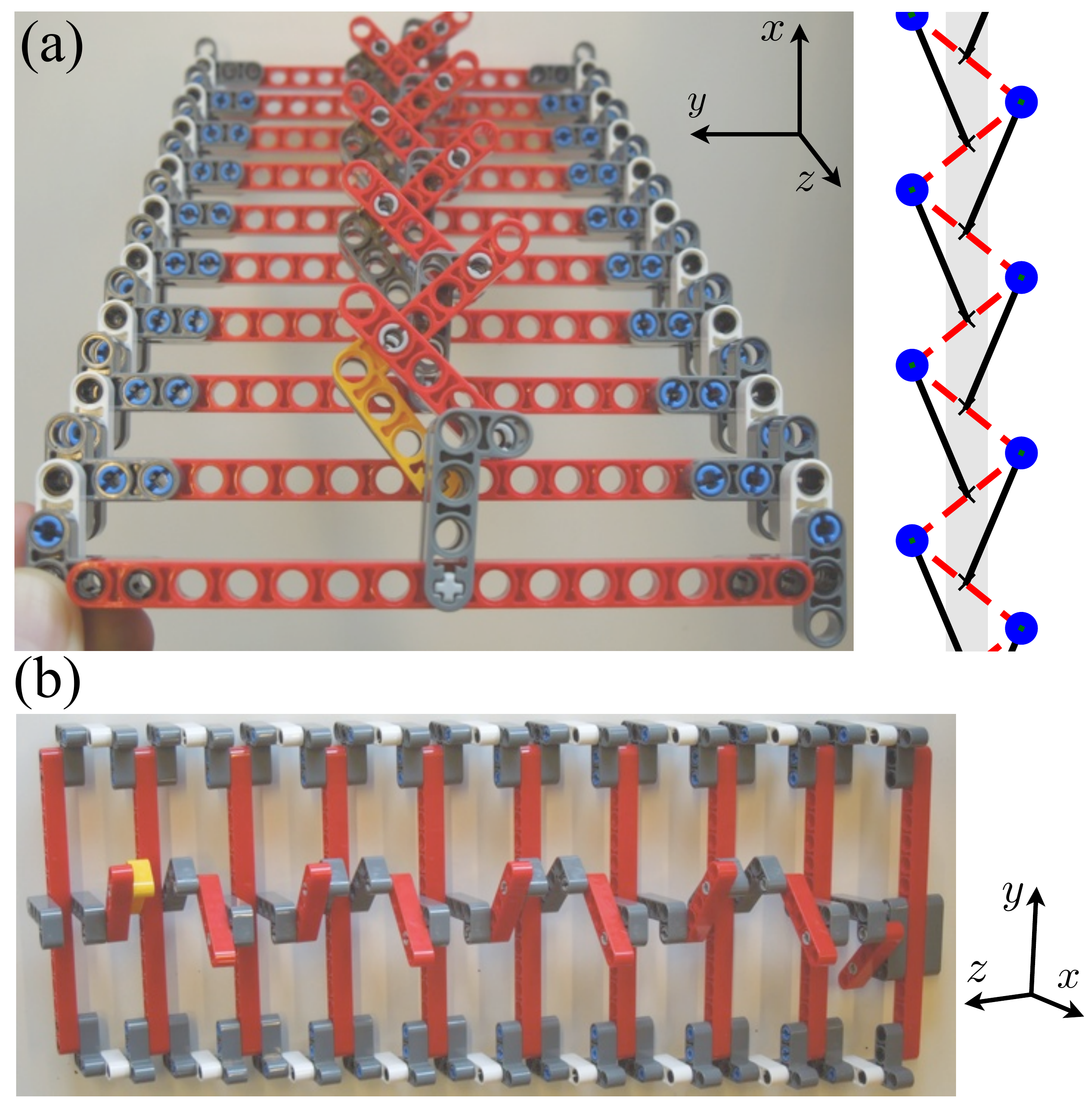}
  \caption{\label{fig:spinnerphoto} A LEGO realization of a chain admitting a spinner soliton (see also Movie S9)
(a) A view of the realization alongside a computer
graphics sketch.  The ``L''-shaped pieces are the rotors, the shorter
red bars connecting their ends are the springs and the long red bars
serve as supports for the pinning. The ``T''-shaped pieces connect the
red bars to a single rigid background frame. (b) Another view showing the
staggering of the rotors in the transverse ($z$) direction.}
\end{figure}

In order to build a spinner prototype, the 3D
embedding must avoid the many self-intersections visible in Fig.\
(\ref{fig:fig3}).
A model made from LEGO pieces is shown in
Fig.\ (\ref{fig:spinnerphoto}) and Movie S9. 
The crucial point is that the rotors are staggered 
in the {\it transverse} direction, causing the transverse width to be
proportional to the number of rotors, see the $z-$axis in Fig.\
(\ref{fig:spinnerphoto}).  

To sum up, the zero-energy moving domain walls in simple model chains
are basic examples of a generic physical process that we expect to exist in more complex man-made and natural structures relevant to robotics and mechanical metamaterials. The protected excitations of topological mechanical metamaterials, that appear as zero-energy vibrational modes within the linear approximation may integrate to finite deformations capable of transporting a mechanical state across the system. Our work raises the question of whether the principle of topological robustness can be adopted in the design of robotic manipulators composed of activated linkages as well as in the fabrication of complex molecular nanostructures.




\begin{acknowledgments}
We acknowledge financial support from FOM and NWO. We thank J.C.Y.\
Teo, A.\ Turner, J.\ Paulose and Y.\ Zhou for helpful discussions and are grateful to
T.-s. Chen for assistance in constructing the flipper.
\end{acknowledgments}





\newpage

\part*{Supporting information: Appendix}

\section{Soliton in the flipper phase}

\setcounter{equation}{0}
\renewcommand\theequation{S\arabic{equation}}

\setcounter{figure}{0}
\renewcommand\thefigure{S\arabic{figure}}

In this section, we derive the continuum $\phi^4$-theory that
describes the domain wall (soliton) solution in the flipper phase of
motion. The discrete Lagrangian for the one-dimensional mechanical
analog of the SSH chain is:
\begin{align}
L &= \sum _n \frac{1}{2}I\left(\frac{d\theta_n}{dt}\right)^2 -
\frac{1}{2}k_e\left(l_{n,n+1}-\bar{l}\right)^2,\label{discrete_Lagrangian}
\end{align}
where, $ I=Mr^2 $ is the moment of inertia of the rod-mass system,
$\theta _n$ is the angular position of the $n$th rod measured with respect to the positive direction of the $y$- axis, $k_e$ is the bare spring constant between neighbouring rods with rest length $\bar{l}$ and $l_{n,n+1}$ is the instantaneous length of the spring connecting rods $n$ and $n+1$.

We make the working assumption that throughout the course of the
motion, the lengths of the springs do not change appreciably from
their rest length, that is, we will consider the springs to be almost rigid. Thus, a more tractable continuum limit follows by expressing the potential energy in Eq.\ (\ref{discrete_Lagrangian}) as
\begin{align}
\frac{1}{2}k_e\left(l_{n,n+1}-\bar{l}\right)^2 =& \frac{1}{2}\frac{k_e}{(l_{n,n+1}+\bar l)^2}(l_{n,n+1}+\bar l)^2(l_{n,n+1}-\bar l)^2\\
\approx& \frac{\kappa}{2}(l^2_{n,n+1}-\bar l^2)^2,
\end{align}
where in the last step we have assumed that $(l_{n,n+1}+\bar
l)^2\approx 4\bar l^2$ and so the potential energy is approximately
unchanged if we define a new spring constant $\kappa =
\dfrac{k_e}{4\bar{l}^2}$.  Next, by expressing $l_{n,n+1}$ in terms of
the angles $\theta_n,\theta _{n+1}$ and the geometrical parameters
$r,a$, we find the potential part of the Lagrangian to be
\begin{align}
V_{n,n+1}[\theta_n,\theta_{n+1}]=& 2\kappa r^4\left[\cos(\theta _n +
\theta _{n+1}) -  \cos(2\bar\theta)\right.\\
&\left. + \frac{a}{r}(\sin\theta _n -
\sin\theta_{n+1})\right]^2.\nonumber
\end{align}
Next, we take the continuum limit, assuming
$a\rightarrow dx$, $\theta
_n\rightarrow\theta(x)-\frac{a}{2}\frac{d\theta}{dx}$, $ \theta _{n+1}
\rightarrow \pi- \theta(x)-\frac{a}{2}\frac{d\theta}{dx}$ (taking a
Taylor series centered at $x=n+1/2$), to obtain the
potential energy density as a perfect square 
\begin{align} 
V[u(x)]=&\frac{K}{2}\left[(\bar
u^2-u(x)^2)+\frac{a^2}{2}\frac{du(x)}{dx}\right]^2,\label{eq:potentialtermperfectsquare}
\end{align}
where we denote the projection of the rods along the $\hat{x}-$axis by
$u(x)=r\sin\theta(x)$ and define the constants $\bar
u=r\sin\bar\theta$ and $K=16\kappa=\dfrac{4k_e}{\bar l^2}$. 

Thus, in the continuum limit and expressed in terms of the
field $u(x,t)$, Eq.\ (\ref{discrete_Lagrangian}) assumes the form:
\begin{align}
\mathcal{L}=&\int \frac{1}{2}\frac{Mr^2}{r^2-u^2}\left(\frac{\partial
u}{\partial t}\right)^2-\frac{1}{2}K\frac{a^4}{4}\left(\frac{\partial
u}{\partial x}\right)^2 \label{slagrangian2} \\
&-\frac{1}{2}K(\bar u^2-u^2)^2-\frac{1}{2}K\frac{a^2}{2}(\bar
u^2-u^2)\frac{\partial u}{\partial x}.\nonumber
\end{align}
Eq.\ (\ref{slagrangian2}) differs from the ordinary $\phi^4$ theory due
to the presence of the last term (linear in $\frac{\partial
u}{\partial x}$) and because of the non-linear kinetic term. As explained
in the main text, the last term is related to the topological charge
of the soliton that ensures that the static kink costs zero potential
energy and does not contribute to the equation of motion. Moreover,
by expanding the nonlinear kinetic term to order $u^2/r^2$ (valid in
the limit of small $\bar{\theta}$), we obtain from Eq.\
(\ref{slagrangian2}) the ordinary $\phi^4$ theory, whose soliton
solution is 
\begin{eqnarray}
u = \bar u\tanh\left[\frac{x-x_0-vt}{\frac{a^2}{2\bar
u}\sqrt{1-\frac{v^2}{c^2}}}\right].\label{SKink2}
\end{eqnarray}
where $v$ is the speed at which the kink propagates and the effective
``speed of light'' is $c=\dfrac{a^2}{\bar l}\sqrt{\frac{k_e}{M}}$. This
coincides with Eq.\ (11) in the main text.


%
%
%

\section{Soliton in the spinner phase}
As discussed in the main text, in order to obtain the soliton solution in the spinner phase, we define the continuum field as the slowly varying angular field of odd or even sites. In the following, we discuss an approximate method for describing odd (even) fields,
whose soliton solutions can be described by the sine-Gordon theory. Consider Eq.\ (\ref{discrete_Lagrangian}) (writing three terms),
assuming $n$ is an odd site for instance and with the working assumption that $(l_{n,n+1}+\bar l)^2\approx 4\bar l^2$:
\begin{align}
L =& \sum_n \frac{1}{2}I\dot{\theta}^2_n +
\frac{1}{2}I\dot{\theta}^2_{n+1} + \frac{1}{2}I\dot{\theta}^2_{n+2}\\
& - \frac{k_e}{8\bar{l}^2}(l^2_{n,n+1}-\bar{l}^2)^2-\frac{k_e}{8\bar{l}^2}(l^2_{n+1,n+2}-\bar{l}^2)^2\nonumber\\
& -\frac{k_e}{8\bar{l}^2}(l^2_{n+2,n+3}-\bar{l}^2)^2,\nonumber
\end{align}
where dots denote derivative with respect to time. 

In order to combine odd (even) sites, consider combining half of the potential energies at a time, so that both odd and even sites get
their share from the same Lagrangian. To do this, we express the potential energy as sum of terms of the following form:
\begin{align}
V =& \frac{k_e}{16\bar{l}^2} \sum_n\left\{ \left[(l^2_{n,n+1}-\bar{l}^2)^2 +
(l^2_{n+1,n+2}-\bar{l}^2)^2\right]\right.\\
& \left. +\sum_n \left[(l^2_{n+1,n+2}-\bar{l}^2)^2 +
(l^2_{n+2,n+3}-\bar{l}^2)^2\right]\right\}.\nonumber
\end{align}
For instance, the first square bracket can now be used to integrate out an even site and the second one to integrate out an odd site. Consider the first square bracket re-expressed as
\begin{align}
\left[(l^2_{n,n+1}-\bar{l}^2)^2 + (l^2_{n+1,n+2}-\bar{l}^2)^2\right]
=&  \frac{1}{2}\left[(l^2_{n,n+1} + l^2_{n+1,n+2}
-2\bar{l}^2)^2\right.\nonumber\\
&\left. + (l^2_{n,n+1}-l^2_{n+1,n+2})^2\right].
\end{align}
Assuming that the average of $l^2_{n,n+1} + l^2_{n+1,n+2}=2\bar{l}^2$, the first term in the above equation can be approximated to 0.  
We are thus left with 
\begin{align}
\left[(l^2_{n,n+1}-\bar{l}^2)^2 + (l^2_{n+1,n+2}-\bar{l}^2)^2\right]=
\frac{1}{2}\left[ (l^2_{n,n+1}-l^2_{n+1,n+2})^2\right].
\end{align}
After substituting the lengths by angles and $r,a$, we obtain 
\begin{align}
l^2_{n,n+1}-l^2_{n+1,n+2}=&2r^2\left[\cos(\theta _n-\theta
_{n+1})-\cos(\theta _{n+1}-\theta _{n+2})\right.\nonumber\\
&\left.+\frac{a}{r}[\sin(\theta_n)+\sin(\theta _{n+2}) - 2\sin(\theta
_{n+1})]\right].
\end{align}
As discussed in the main text, we take the continuum limit by defining the field $\theta _n\rightarrow \theta (x)$ and $\theta
_{n+2}\rightarrow \theta(x+2a)=\theta + 2a\frac{d\theta}{dx}$, and then retaining terms to leading order in $a$. We ``integrate out'' $\theta_{n+1}$ (the degree of freedom representing the middle rod) by using the constraint equation $l^2_{n,n+1}=\bar{l}^2$. Expressing this second constraint equation in terms of the angles $\{ \bar{\theta}, \theta _{n}, \theta _{n+1}\}$, we find (in the limit $a \ll r$) that $\theta _{n+1} \approx \theta _{n} + \pi -2\bar{\theta}$. We thus obtain the following potential energy contribution from the odd sites:
\begin{align}
V_{o} =&\frac{k_er^4a^2\sin^2(2\bar{\theta})}{2\bar{l}^2}\left[\frac{d\theta}{dx} - \frac{a}{r\sin\bar{\theta}}\sin(\theta-\bar{\theta})\right]^2.
\end{align}
We therefore identify an effective spring constant $K^{\text{eff}}= \dfrac{k_er^4a^2\sin^2(2\bar{\theta})}{\bar{l}^2}$. Similarly, combining half of the kinetic energies from the the odd sites $\frac{1}{4}Mr^2\dot{\theta}^2_n+\frac{1}{4}Mr^2\dot{\theta}^2_{n+2}$, we obtain $\frac{1}{2}Mr^2\dot{\theta}^2$. Therefore, the moment of intertia is $I=Mr^2$. 

With this procedure, we have therefore expressed the Lagrangian as a
sine-Gordon Lagrangian for the odd (even) sites, whose continuum limit for the odd (o) sites reads:
\begin{align}
\mathcal{L}_o =& \int dx
\left\{\frac{1}{2}I\dot{\theta^2_{o}}-\frac{1}{2}K^{\text{eff}}\left[\theta'_{o}
-\frac{a}{r\sin\bar{\theta}}\sin(\theta
_{o}-\bar{\theta})\right]^2\right\} ,
\end{align}
while for the even (e) sites, we find
\begin{align}
\mathcal{L}_e =&  \int dx\left\{\frac{1}{2}I\dot{\theta^2_{e}}-
\frac{1}{2}K^{\text{eff}}\left[\theta'_{e} +
\frac{a}{r\sin\bar{\theta}}\sin(\theta
_{e}+\bar{\theta})\right]^2\right\},
\end{align}
where primes denote derivative with respect to space $x$. 
Upon using the Euler-Lagrange equations, we obtain the respective soliton solutions:
\begin{align}
\cos(\theta\pm\bar{\theta}) =&\pm \tanh\left(\frac{x-vt}{r\sin(\bar{\theta})\sqrt{1-\frac{v^2}{c^2}}}\right),
\label{Ssgd}
\end{align}
where, $\pm$ correspond to soliton solutions for the even (odd) sites
respectively (which are of course constrained so that
$\theta_e=\theta_o+\pi-2\bar\theta$) and where $c$, the speed of sound in the spinner phase $r\gg a$ is  
\begin{align}
c= \sqrt{\frac{K^{\text{eff}}}{I}}=&\frac{ar\sin(2\bar\theta)}{\bar{l}}\sqrt{\frac{k_e}{M}}.
\end{align}

The sine-Gordon soliton solution in the spinner phase suggests an
analogy with the well known mechanical model consisting of a chain of
pendula coupled via torsional springs \cite{Sdauxois}. However, a side
view of the model spinner chain in Fig.\ (7b) suggests several
important differences between the two models.  First, unlike for the
chain of coupled pendula, the springs connecting neighbouring rods in
our model have non-zero equilibrium projection in the plane of
rotation of the rods and thus, the even- and odd-numbered sites have
different equilibrium positions. Second, the rotors do not all rotate
about the same axis, but adjacent axes are displaced by the lattice
spacing $a$ in the $x-$direction, see Fig.\ (7) in the main text.  This breaks the global rotational symmetry of the chain. Moreover, as for the flipper phase, the sine-Gordon soliton has a non-zero topological charge originating from the term linear in $\theta '$. This charge ensures that the static kink has zero energy, while in the dynamical case, it lowers the total energy of the soliton by a constant factor. 


\section{Linearized perturbation theory for the wobbling flipper}
We now construct a simple model to qualitatively understand the
wobbling flipper phase of motion as a superposition of the $\phi^4$
flipper kink and linear perturbations around its asymptotic states. Since the wobbling flipper phase is observed for $d > 1$ (where $d=\frac{2r\sin\bar{\theta}}{a}$), we find that to correctly account for the spatial period and decay length of the oscillations around the flipper kink, the linear theory must bear signatures of the spinner phase of motion, see Eq.\ (4) and the following discussion in the main text.

\begin{figure}
 \centering
 \includegraphics[width=.44\textwidth]{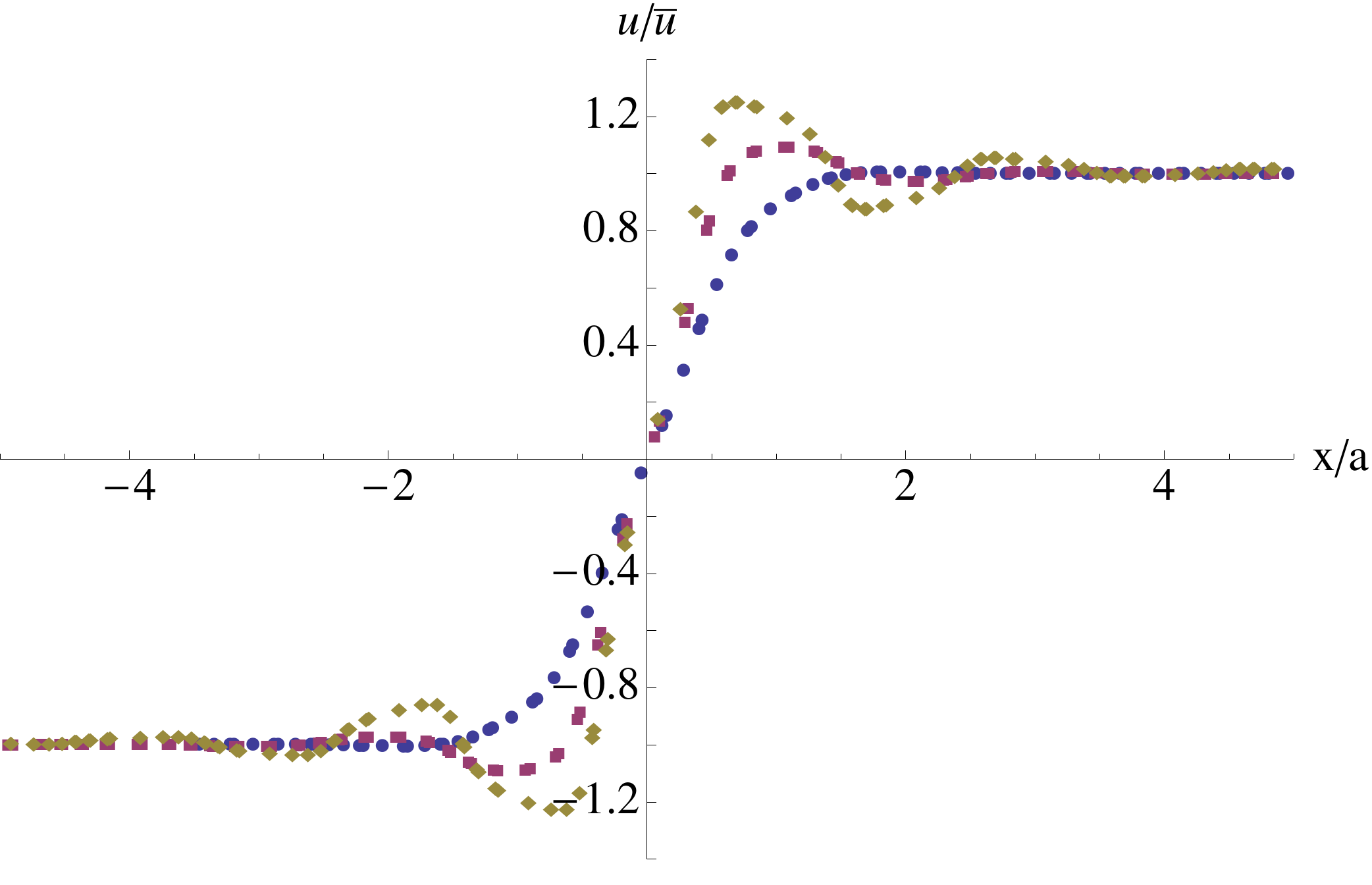}
 \caption{\label{fig:appendixfig1} The numerically
generated displacement field $u(x-vt)$ (symbols) for various $r/a$ in the wobbling flipper phase. 
These results are for a chain of 100 rods with
$\bar\theta=0.77$ and $k_e=1,M=1$, with $r/a=0.8,d=1.1$ (blue),
$r/a=1.2,d=1.7$ (magenta)
and $r/a=1.6,d=2.2$ (gold). The data
arise from 10 snapshots of a single trajectory, translated so that
the center of the soliton is at $x=0$. }
\end{figure}

\begin{figure}
 \centering
 \includegraphics[width=.44\textwidth]{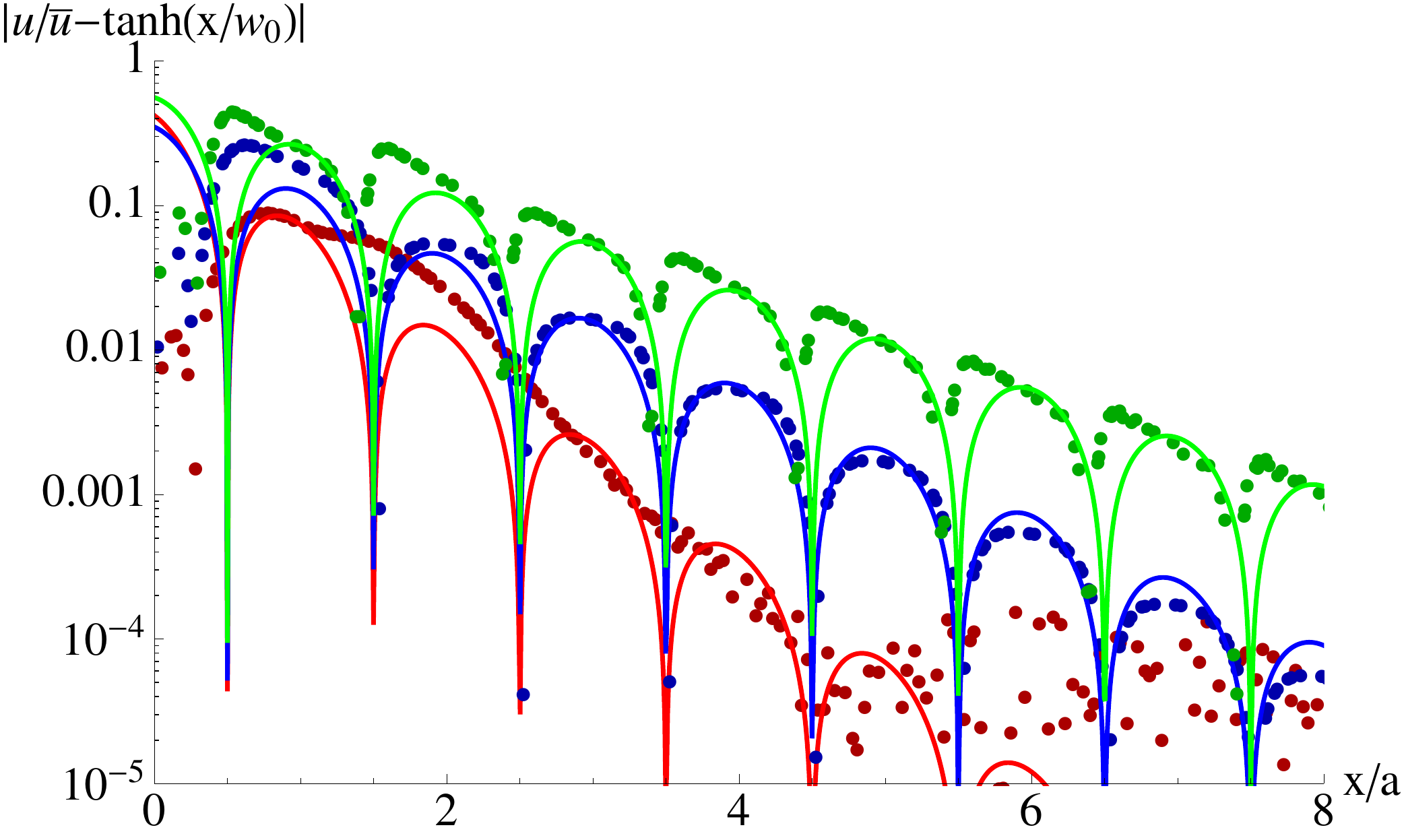}
 \caption{\label{fig:appendixfig2} The oscillations in numerically
generated displacement fields $u(x-vt)$ (symbols) in the wobbling
flipper phase compared with the functions $e^{-x/w_s}\cos(\pi x/a)$
(solid lines).  Oscillations are quantified by subtracting off the hyperbolic tangent flipper solution.
The numerical results are for a chain of 100 rods with
$\bar\theta=0.77$ and $k_e=1,M=1$, with $r/a=0.8,d=1.1$ (red),
$r/a=1.3,d=1.8$ (blue)
and $r/a=2.0,d=2.8$ (green). The solid lines have a decay length of
$w_s/a=(r/a)\sin\bar\theta=d/2$ with the same values of $r/a,d$. The data
arise from 10 snapshots of a single trajectory, translated so that
the center of the soliton is at $x=0$.}
\end{figure}

We show the main phenomena as we increase $d$ in the wobbling flipper
phase in Figs.\ (\ref{fig:appendixfig1}) and (\ref{fig:appendixfig2}).
The static profile of the wobbling flipper is distinguished from that
of the non-wobbling flipper by oscillations in $u$ around the value $\bar u$.
For small $d-1$, the profile is indistinguishable by eye from the
hyperbolic tangent profile of the non-wobbling flipper, as the
amplitude of the oscillations is small, though subtracting off that
profile as a background makes the deviation visible. As $d$ increases, the
amplitude increases; we found that the dependence of the maximum
deviation above the equilibrium value $u=\bar u$ was proportional to
$d^3$. Furthermore, the wavelength of the oscillations seems to
stabilize at $2a$. The function $e^{-x/w_s}\cos(\pi x/a)$ was found to fit the
oscillations for large enough $x/a$ and $d$.  

In the following, we approximate the oscillations observed in the
wobbling phase (see the top inset of Fig.\ (5) top inset in the main
text) as small perturbations around the uniform ground state
$\theta=\pm\bar{\theta}$ that the kink profiles approach
asymptotically. 
Since the width of the flipper kink decreases inversely as $w\sim d^{-1}$, the asymptotic values are reached within a few lattice spacings for larger values of $d$. 

We begin with Eq.\ (5) in the main text, the equation constraining two adjacent rotor angles
$\theta_n,\theta_{n+1}$. 
\begin{align}
\cos(\theta _{n} + \theta _{n+1}) - \cos(2\bar{\theta}) +
\frac{a}{r}(\sin\theta _{n} - \sin\theta _{n+1})&=0.
\label{linearizeddiscrete}
\end{align}

However, now we keep a few more terms in the continuum limit, again by
taking a Taylor series around $x=n+1/2$:
$\theta_n\rightarrow \theta(n-1/2) \approx
\theta(x)-\frac{a}{2}\frac{d\theta}{dx}+\frac{a^2}{8}\frac{d^2\theta}{dx^2}$,
$\theta_{n+1}\rightarrow \theta(n+1/2) \approx
\theta(x)+\frac{a}{2}\frac{d\theta}{dx}+\frac{a^2}{8}\frac{d^2\theta}{dx^2}$.
To this order, we obtain
\begin{align}
\cos2\left(\theta + \frac{a^2}{8}\frac{d^2\theta}{dx^2}\right) - \cos(2\bar{\theta}) +
\frac{a^2}{r}\left(\frac{d(\sin\theta)}{dx}\right)&=0.
\end{align}

Observe that this continuum limit respects the reflection symmetry
$(x,\theta)\rightarrow (-x,-\theta)$.

It is convenient to define
$\theta^*=\theta+\frac{a^2}{8}\frac{d^2\theta}{dx^2}$. Note that
$\cos(2x)-\cos(2y)=(\sin x+\sin y)(\sin y-\sin x)$.  We use this fact
and then (after multiplying through by $r^2$) use the approximation $\sin\bar\theta+\sin\theta^*\approx2\sin\bar\theta$ 
as we'd like to consider the behavior when $\theta\approx\bar\theta$
\begin{align}
2r^2(2\sin\bar\theta)(\sin\theta^*-\sin\bar\theta) +
a^2r\left(\frac{d(\sin\theta)}{dx}\right)&=0.
\end{align}

We assume
that $\epsilon=\frac{a^2}{8}\frac{d^2\theta}{dx^2}$ is small enough
that $\sin\epsilon\approx \epsilon$ and $\cos\epsilon\approx 1$.  Then
\begin{align}
2r^2(2\sin\bar\theta)\left(\sin\theta-\sin\bar\theta+\frac{a^2\cos\theta}{8}\frac{d^2\theta}{dx^2}\right) +
a^2r\left(\frac{d(\sin\theta)}{dx}\right)&=0.
\end{align}

As usual, let $u=r\sin\theta$ and $\bar u=r\sin\bar\theta$. A
consequence of our assumptions is that $\frac{d^2u}{dx^2}\approx
r^2\cos\theta\frac{d^2\theta}{dx^2}$ (neglecting a term nonlinear in
$u$). Then
\begin{align}
4\bar u\left(u-\bar u+\frac{a^2}{8}\frac{d^2u}{dx^2}\right) +
a^2\frac{du}{dx}&=0\\
\frac{a^2}{8}\frac{d^2(\delta u)}{dx^2} +
\frac{a^2}{4\bar u}\frac{d(\delta u)}{dx}+\delta
u&=0.\label{wobblingode}
\end{align}

In the last line, we introduce $\delta u\equiv u-\bar u$.  The
solutions to Eq.\ (\ref{wobblingode}) are linear combinations of
complex exponentials.  In particular, 
\begin{align}
\delta
u&=Ae^{\lambda_-x}+Be^{\lambda_+x},
\end{align}
where
\begin{align}
\lambda_\pm&=-\frac{1}{\bar u}\pm\sqrt{\frac{1}{\bar
u^2}-\frac{8}{a^2}}.
\end{align}

Provided $\frac{1}{\bar u}<\frac{\sqrt{8}}{a}$ or $d>1/\sqrt{2}$,
$\lambda_\pm$ are a complex conjugate pair, with real parts equal to $\frac{1}{\bar u}$
(as observed in Fig.\ (\ref{fig:appendixfig1}) and Fig.\
(\ref{fig:appendixfig2})) and imaginary parts approaching $\pm\sqrt{8}i/a\approx
2.8ia$ (compared to the wavenumber of $\pi i/a\approx 3.1 i/a$ that we
observe).  Thus for sufficiently large $d$, this linearized perturbation theory captures the fact
that the decay length of the oscillations is equal to $\bar
u=r\sin\bar\theta$, the width of the spinner soliton, and the
wavenumber is on the order of the lattice spacing. In the opposite
limit ($d<1/\sqrt{2}$), both roots are real negative and thus the
solution is simply an exponential decay, agreeing qualitatively with
the shape of the tail of the hyperbolic tangent kink solution in the
non-wobbling flipper phase.

We conjecture that with a more sophisticated expansion (e.g.\ with
higher order terms or some treatment of the nonlinear effects), the
value of $\bar u/a$ and $d$ at which we begin to see oscillations (and
hence which represents the transition between non-wobbling and wobling
flippers) should approach 1, and the imaginary part should
approach $\pi i/a$.

\end{article}
\end{document}